\documentclass{revtex4}

\usepackage{epsf}
\usepackage[dvips]{color}

\newcommand{\expect}[2]{\big< \, #1 \, \big| \,#2\, \big|\, #1 \,\big>}
\newcommand{\op}[1]{\tilde{#1}} 

\begin{document}

\title{Two-neutron correlations in 
microscopic wave functions of $^6$He, $^8$He and $^{12}$C}

\author{Yoshiko Kanada-En'yo}
\affiliation{Department of Physics, Kyoto University,
Kyoto 606-8502, Japan}

\author{Hans Feldmeier}
\affiliation{Gesellschaft f\"ur Schwerionenforschung, Planckstra\ss e 1, 64291 Darmstadt, Germany}

\author{Tadahiro Suhara}
\affiliation{Yukawa Institute for Theoretical Physics, Kyoto University,
Kyoto 606-8502, Japan}

\begin{abstract}
Two-neutron densities obtained from microscopic wave functions of
$^6$He and $^8$He are investigated 
to reveal di-neutron correlations.
In particular, the comparison of the two-neutron density with 
the product of one-neutron densities is 
useful for a quantitative discussion of di-neutron correlations.
The calculations show that 
the $S=0$ spatial two-neutron correlation increases
at the surface of $^6$He$(0^+_1)$ and $^8$He$(0^+_2)$. The enhancement 
is remarkable in the $^6$He$(0^+_1)$ ground state but not as prominent in the  
$^8$He$(0^+_1)$ ground state.
Configuration mixing of many Slater determinants is 
essential to describe the di-neutron correlations.
Two-neutron densities in $^{12}$C wave functions with $\alpha$-cluster structures 
are also studied.
\end{abstract}

\maketitle

\noindent

\section{Introduction}

Two-neutron ($nn$) correlations in neutron-rich nuclei are presently getting a lot of attention 
because the nuclear force  shows a rather strong attraction in the $^1S_0$ channel 
which, however, is not strong enough to form a two-neutron bound state,
instead there is a di-neutron scattering resonance at low energy. 
Thus neutrons are an example of the general problem how a low energy many-body system 
can be understood microscopically when  
the scattering length is larger than the extension of the many-body system.
Interesting questions are:
what kind of correlations are induced by the interaction and
can one identify generic properties of loosely bound neutrons.

The experimental area of operation are weakly bound neutron-rich nuclei.
For example, in neutron halo nuclei such as $^{11}$Li \cite{tanihata85},
a spatially correlated two-neutron pair outside a core was theoretically 
predicted in many works
\cite{Hansen:1987mc,Bertsch:1991zz,Zhukov:1993aw,Esbensen:1997zz,ikeda02,aoyama01,Myo:2002wq,
Myo:2003bh,Hagino:2005we,myo07} 
and was supported by experiments
\cite{ieki93,sackett93,shimoura95,zinser97,nakamura06}.

$nn$ correlations have been intensively investigated for asymmetric nuclear matter 
\cite{bardo90,Takatsuka:1992ga,DeBlasio:1997zz,Dean:2002zx,Baldo:2005qi,
 Matsuo:2005vf,Margueron:2007uk,Isayev:2008wu},
and are  discussed also in light neutron-rich nuclei such as $^8$He
\cite{KanadaEn'yo:2007ie,Itagaki:2008zz,Hagino:2008vm,fmd05,fmd04b,fmd08} and 
in medium-heavy neutron-rich nuclei \cite{Matsuo:2004pr,Pillet:2007hb}. 
These studies suggest that the $nn$ correlations play an enhanced role 
at the nuclear surface of finite nuclei
and in model studies of infinite neutron matter at low density (weak interactions turned off).

As the neutron-neutron interaction forms at low energies a localized resonance 
in the $l=0$ channel, 
it is expected that spatial correlations between two neutrons that are loosely
bound to a core are enhanced compared to well bound nuclear many-body systems
where they are in close contact to protons. 
In two-neutron halo nuclei, the $nn$ correlations between valence neutrons
are for example investigated by means of semi-microscopic three-body models 
describing the center of mass (c.m.) motion of a core and two neutrons. 
The antisymmetrization between a valence neutron and the core is not taken into 
account microscopically but it is treated semi-microscopically by 
considering Pauli forbidden orbits for the valence neutron motion.
In such semi-microscopic models, two of the neutrons are regarded 
as valence neutrons and distinguished from the others. 
As their positions are the only degrees of freedom,
one is able to discuss the $nn$ correlations by simply analyzing 
the relative motion between the two valence neutrons.
In reality, however, neutrons are indistinguishable fermions and 
the nuclear many-body system has to be expressed by 
fully antisymmetrized wave functions \cite{fmd90,FelSchn00,ENYObc}. 
Therefore it is difficult to discern, at least in the inner region, valence neutrons 
from core neutrons.

%

Generally speaking, in quantum many-body systems 
consisting of indistinguishable particles, observable quantities  
are represented by hermitian operators that are symmetric under particle exchange.
The two-body density satisfies this demand
and contains all two-body information about the system. Thus it represents
the basis for studies of $nn$ correlations in theoretical frameworks 
and experimental observations.

Here we should remind the reader that microscopic wave functions always contain 
correlations that one regards as trivial but are also reflected in the two-body density.
Firstly, antisymmetrization or the Pauli principle has significant effects on the 
two-body correlations. For example, if a spin-zero $nn$ pair is chosen, only the 
parity-even state is allowed but the spatially odd components are forbidden in 
the relative wave function. 
Secondly, energy eigenstates of a finite nuclear system are also eigenstates of
parity and total-angular-momentum. 
Therefore intrinsic states have to be projected onto parity and 
total-angular-momentum eigenstates. This implies that even an intrinsic
state without specific correlations leads to a many-body state that contains
long-ranged many-body correlations.
Thirdly, in case of an even-even nucleus, where in a shell-model picture with independent 
particles a single-particle $j$-shell is not completely filled, 
angular momentum coupling to $J^\pi=0^+$ of the ground state leads to
two-body correlations. 
Already these three examples show that
it is essential to distinguish carefully trivial correlations from non-trivial
ones that are induced by the nuclear interaction beyond the mean-field level.

One possible definition of two-body correlations is to take the difference of 
the two-body density calculated with the correlated many-body state 
minus the antisymmetrized product of its one-body densities so that
the trivial part from the Pauli principle is not regarded as a correlation.
This definition is in the spirit of classical probability theory where correlations 
between two random variables are defined by the difference between the 
joined distribution and the product of the reduced distributions (Sec. \ref{sec:prod}).
Another possibility is based on the independent particle mean-field picture.
The one-body density of the correlated many-body state can be used to calculate
from the Hamiltonian a one-body mean-field Hamiltonian whose lowest eigenstate
is a single Slater determinant which then may be used as the uncorrelated reference state.
However, this Slater determinant is spherical for a $J^\pi = 0^+$ state and is not
the Hartree-Fock ground state.
This discussion shows that a general nuclear many-body wave function contains
various kinds of many-body correlations and there is no  unique and straightforward
definition of correlations. 

The aim of this paper is to study $nn$ correlations  
by analyzing microscopic many-body wave functions
obtained by means of antisymmetrized molecular dynamics (AMD) \cite{ENYObc,KanadaEn'yo:2006ze,ENYOsup,AMDrev}.
In previous work AMD calculations described the structures of $^6$He and $^8$He in
Ref.~\cite{KanadaEn'yo:2007ie} quite well.
The AMD wave functions are linear combinations of many Slater determinants and thus
incorporate various types of correlations. 
An important advantage of the AMD wave functions is that
the c.m. motion is completely decoupled from the intrinsic one.

After defining in Sec. \ref{sec:formulation} the one- and two-body densities that will be used
and explaining the AMD many-body states which are a superposition of many 
angular momentum and spin projected many-body Slater determinants
we calculate the two-neutron densities of these correlated wave functions and discuss
the $nn$ correlations in the ground states of $^6$He and $^8$He in Sec.~\ref{sec:he-isotopes}.
To clarify the enhancement of the $nn$ correlations, we prepare reference wave functions 
in form of Slater determinants which are regarded to possess only trivial correlations
and in this sense are uncorrelated.
Then we show characteristic features of two-neutron densities calculated with AMD wave functions 
in comparison with the results of uncorrelated wave functions.
We also perform a similar analysis for $^{12}$C wave functions and discuss in Sec.~\ref{sec:12C}
how $\alpha$-cluster structures show up in the two-neutron density.
Finally, in Sec.~\ref{sec:summary} we summarize and give an outlook. 

\section{Formulation}\label{sec:formulation}
\subsection{One-body and two-body density}

Let us consider an antisymmetrized many-body wave function $\Phi$ 
which represents an $A$-nucleon system.
The one-body density is defined as 
\begin{equation}
\rho^{(1)}({\bf R})=\expect{\Phi}{\sum_{i=1}^A
\delta(\tilde{\bf r}_i-{\bf R})} \ .
\end{equation}
Here $\tilde{\bf r}_i$ is the position operator for the $i$th particle
and $\Phi$ is normalized to one. 
Using the operator $\tilde{\tau}_z$ for the $z$-component of the Pauli matrices in isospin space
the density can be decomposed into neutron and proton density.
\begin{equation}
\rho^{(1)}_{\{ {n \atop p} \}} ({\bf R})=
\expect{\Phi}{\sum_{i=1}^A \left(\frac{1\mp \tilde{\tau}_{zi}}{2}\right)
 \delta(\tilde{\bf r}_{i}-{\bf R}) } \ .
\end{equation}
%
%
In a similar way the two-body density is defined as 
\begin{equation}
\rho^{(2)}({\bf r}_1,{\bf r}_2)=
\expect{\Phi}{\sum_{i\ne j}^A
\delta(\tilde{\bf r}_i-{\bf r}_1)\ \delta(\tilde{\bf r}_j-{\bf r}_2)} \ ,
\end{equation}
where $\rho^{(2)}({\bf r}_1,{\bf r}_2)$ is the probability density that one nucleon
is found at the position ${\bf r}_1$ and another one at ${\bf r}_2$.
The two-body density can be rewritten in terms of relative 
${\bf r}\equiv {\bf r}_2-{\bf r}_1$ 
and c.m. position ${\bf R}\equiv ({\bf r}_1+{\bf r}_2)/2 $,
\begin{equation}
\rho^{(2)}({\bf r}_1,{\bf r}_2)=\rho^{(2)}({\bf R},{\bf r})=
\expect{\Phi}{\sum_{i\ne j}^A
\delta(\tilde{\bf R}_{ij}-{\bf R})\ \delta(\tilde{\bf r}_{ij}-{\bf r}) } \ ,
\end{equation}
where $\tilde{\bf r}_{ij}\equiv\tilde{\bf r}_j-\tilde{\bf r}_i$ and 
$\tilde{\bf R}_{ij}\equiv(\tilde{\bf r}_j+\tilde{\bf r}_i)/2$.
The two-body density for neutron and proton pairs is given by
\begin{equation}
\rho^{(2)}_{\{ {nn \atop pp} \}} ({\bf R},{\bf r})=
\expect{\Phi}{\sum_{i\ne j}^A \left(\frac{1\mp \tilde{\tau}_{zi}}{2}\right)
                       \left(\frac{1\mp\tilde{\tau}_{zj}}{2}\right)
\delta(\tilde{\bf R}_{ij}-{\bf R})\   \delta(\tilde{\bf r}_{ij}-{\bf r})}.
\end{equation}
The two-neutron density $\rho^{(2)}_{nn}$ (and likewise the one for protons)
can be separated into densities for
spin zero and for spin one pairs 
\begin{equation}
\rho^{(2)}_{nn}({\bf R},{\bf r})=\rho^{(2)}_{nn, S=0}({\bf R},{\bf r})+
\rho^{(2)}_{nn,S=1}({\bf R},{\bf r}) \ ,
\end{equation}
with
\begin{equation}
\rho^{(2)}_{nn,S=\{ {0 \atop 1} \}}({\bf R},{\bf r})=
\expect{\Phi}{ \sum_{i\ne j}^A 
\left(\frac{1-\tilde\tau_{zi}}{2}\right)
\left(\frac{1-\tilde\tau_{zj}}{2}\right)
\left(\frac{1\mp\tilde P^\sigma_{ij}}{2}\right)
\delta(\tilde{\bf R}_{ij}-{\bf R})\ \delta(\tilde{\bf r}_{ij}-{\bf r})
} \ .
\end{equation}
$\tilde P^\sigma_{ij}$ is the spin-exchange operator,
and $(1\mp \tilde P^\sigma_{ij})/2$ projects on $S=0$ and $S=1$, respectively.

In order to reduce the six-dimensional list of arguments we expand
the two-body density in spherical harmonics as
\begin{equation}
\rho^{(2)}({\bf R},{\bf r})=\sum_{LM,lm}
\rho^{(2)}_{LM,lm}(R,r)\ Y_{LM}(\hat {\bf R})\ Y_{lm}(\hat {\bf r})
\end{equation}
and consider
in the present paper only the $l=m=0$ and $L=M=0$ component
which we obtain by 
integrating
over the orientations $\hat {\bf R}$ and $\hat {\bf r}$ as
\begin{equation}\label{eq:9}
\rho^{(2)}(R,r)\equiv \rho^{(2)}_{00,00}(R,r) = \int d\Omega_R \ Y^*_{00}(\hat {\bf R})
\int d\Omega_r \ Y^*_{00}(\hat {\bf r})\ \rho^{(2)}({\bf R},{\bf r}) \ .
\end{equation}
For convenience we omit in the following the subscripts for $L=M=0$ and $l=m=0$.
The lowest components of total two-neutron density, $\rho^{(2)}_{nn}(R,r)$, and its 
$S=0$ and $S=1$ parts, $\rho^{(2)}_{nn,S=0}(R,r)$ and $\rho^{(2)}_{nn,S=1}(R,r)$, 
are calculated in the same way.

The integrated two-body density and two-neutron density equals 
$A(A-1)$ and $N(N-1)$, respectively, 
\begin{eqnarray}
\int d^3r \int d^3R \ \rho^{(2)}({\bf R},{\bf r})=
4\pi \int r^2 dr \int R^2 dR \ \rho^{(2)}(R,r) = A(A-1) \ ,\\
\int d^3r \int d^3R \ \rho^{(2)}_{nn}({\bf R},{\bf r})=
4\pi \int r^2 dr \int R^2 dR \ \rho^{(2)}_{nn}(R,r) = N(N-1) \ ,
\end{eqnarray}
which is twice the number of pairs.

The two-neutron c.m. density for the center of mass of a neutron pair coupled 
to $S=\{0,1\}$ at a distance $R$ from the center of the nucleus is obtained by
integrating $\rho^{(2)}_{nn,S=\{0,1\}}(R,r)$ over the relative distance $r$. 

\begin{equation}\label{eq:rhobar} 
\bar\rho^{(2)}_{nn,S=\{0,1\}}(R)=4\pi\int dr r^2\rho^{(2)}_{nn,S=\{0,1\}}(R,r)\ .
\end{equation}

Here we give a comment on the total center of mass motion.
In fully microscopic wave functions, the 
total c.m. motion is not contained in the wave functions, 
and the above mentioned two-body densities are defined with the coordinate operators, 
$\tilde{\bf r}_{i}$, that measure from the c.m. of the total system.
In AMD wave functions, the total c.m. motion separates from the intrinsic one
when a common width is used for all single-particle Gaussian wave packets. 
In the present work, we eliminate the total
c.m. motion in the calculations of two-body densities.


\subsection{AMD wave functions}
In the following, we analyze the two-body density calculated with AMD many-body 
wave functions for the neutron-rich nuclei $^6$He, $^8$He and for $^{12}$C.
Details of the $^6$He and $^8$He calculations are explained in 
Ref. \cite{KanadaEn'yo:2007ie}, and those $^{12}$C are described in 
Ref. \cite{KanadaEn'yo:2006ze}.
In the framework of AMD, many-body states are represented by
Slater determinants of single-particle Gaussian wave packets, 
\begin{equation}
 \Phi_{\rm AMD}({\bf Z}) =  {{\cal A}} \{  \varphi_1,\varphi_2,...,\varphi_A \},
\end{equation}
where the $i$th single-particle wave function of the $A$-nucleon system 
is written as a product of spatial ($\phi_{\bf X}$), intrinsic spin ($\chi$), and isospin ($\tau$) 
wave functions,
\begin{eqnarray}
 \varphi_i&=& \phi_{{\bf X}i}\ \chi_i\ \tau_i,\\
 \phi_{{\bf X}i}({\bf r}_j) &\propto& 
\exp\bigl\{-\nu({\bf r}_j-\frac{{\bf X}_i}{\sqrt{\nu}})^2\bigr\},
\label{eq:spatial}\\
 \chi_i &=& (\frac{1}{2}+\xi_i)\chi_{\uparrow}
 + (\frac{1}{2}-\xi_i)\chi_{\downarrow}.
\end{eqnarray}
$\phi_{{\bf X}i}$ and $\chi_i$ are characterized by complex variational parameters, 
$ {\bf X}_i\equiv \{
{\rm X}_{1i}$, ${\rm X}_{2i}$, ${\rm X}_{3i}\}$,
and $\xi_{i}$. 
The isospin
function $\tau_i$ is fixed to be up (proton) or down (neutron). 
The width parameter $\nu$ has a common value for each nucleus.
Accordingly, an AMD wave function $\Phi_{\rm AMD}({\bf Z})$
is expressed by the set of variational parameters, ${\bf Z}\equiv 
\{{\bf X}_1,{\bf X}_2,\cdots, {\bf X}_A,\xi_1,\xi_2,\cdots,\xi_A \}$.

The many-body Hilbert space for a given total-angular-momentum $J$ and parity $\pi$
is spanned by a set of linearly independent $J^\pi$-projected AMD states.
Those can be obtained by minimizing
the energy of parity projected AMD states under various constraints. 
Another possibility 
to construct adequate many-body basis states
is to vary the energy with respect to all parameters
contained in ${\bf Z}$ after $J^\pi$ projection.

Solving the many-body eigenvalue problem
\begin{equation}
\op{H}\ \Phi^{J\pi M}_{n} = E^{J \pi}_n\  \Phi^{J\pi M}_{n},
\end{equation}
we obtain the eigenstates of the Hamiltonian $\op{H}$ represented
as a linear combination of the parity and total-angular-momentum 
projected AMD states
\begin{equation} \label{eq:confmix}
\Phi^{J\pi M}_{n} = \sum_{k=1}^{k_{\rm max}}\sum_{K=-J}^J 
                 c^{J\pi,n}_{k,K}\ \op{P}^{J\pi}_{MK}\ \Phi_{\rm AMD}({\bf Z}^{(k)})\ ,
\end{equation}
where $\op{P}^{J\pi}_{MK}$ is the parity and total-angular-momentum projection operator. 
This method is referred to as multiconfiguration AMD. 
For example the number of independent AMD configurations 
for the He-isotopes considered in the next section is $k_{\rm max}=72$.

\subsection{Correlations}
The term ``correlation'' is used to express relations between
entities. As it is utilized quite generally  we want to be
more specific what we mean by correlations among identical nucleons.
For that it is necessary to distinguish between trivial correlations and those
that are characteristic of the system.
For example the relative distance of two nucleons in a nucleus
will be limited by the diameter of the nucleus. This is certainly
a correlation, but it just expresses that the two nucleons belong to the same
nucleus which is trivial.
The probability density to find two identical nucleons at distance zero
vanishes because they are fermions. This correlation we would also like to
consider as trivial.
On the other hand, if for example the relative distance distribution of a spin $S=0$ neutron pair
is more localized than the overall size of the nucleus or if it differs 
in the surface area from the one in the interior then we regard this
correlation as non-trivial.

In general, independent fermions are represented by a single Slater determinant.
Thus an AMD state $\Phi_{\rm AMD}({\bf Z})$ represents $A$ independent
fermions. The only correlation among them is due to the Pauli principle
which we consider as trivial.
Therefore we want to regard the two-body density calculated with a single
Slater determinant 
in general
as uncorrelated.
However, if the minimum energy state $\Phi_{\rm AMD}({\bf Z})$ breaks the 
rotational symmetry of the Hamiltonian and is deformed it contains already 
non-trivial correlations: the nucleons are not distributed equally,
i.e., isotropically around the center of mass.
This deformed single Slater determinant is the intrinsic state and
must not be regarded as an approximation to one of the eigenstates
of the Hamiltonian which have good total angular momentum and parity.

Angular momentum and parity projection restores the symmetries 
and yields states which can be attributed to eigenstates.  As
the projected state is a superposition of 
many Slater determinants, namely the intrinsic state oriented
in all directions given by the three Euler angles,
there is no contradiction in regarding a single Slater determinant as uncorrelated.
In this sense the intrinsically deformed state may already
describe many-body correlations although it is a single Slater determinant. 
In the following we give explicit examples.

A parity projected AMD state (superposition of two Slater determinants), which minimizes the energy, 
has usually lower energy than the minimum energy AMD state. The additional binding energy
is due to correlations which are present in the projected state but not in
the single Slater determinant. The total energy is even lower when the 
minimization is performed after projection on parity and total angular
momentum (VAP). Here more many-body correlations, which are induced by the Hamiltonian,
can be accommodated in the projected state. 
Finally it is obvious that in a configuration mixing calculation, where the 
many-body eigenstates of the Hamiltonian are given by superpositions
of many $J^\pi$-projected Slater determinants (Eq. (\ref{eq:confmix})), one may 
be able to represent
various kinds of correlations 
induced by the
specific nature of the Hamiltonian. 

In order to distinguish trivial from non-trivial or less specific from more specific
correlations, we define many-body reference states and compare their two-body
densities with those of more correlated states. For example a reference
state could be the Hartree-Fock like single AMD Slater determinant.


\section{Results of $^6$He and $^8$He}\label{sec:he-isotopes}
In this section we analyze 
one- and
two-body densities obtained from
$^6$He and $^8$He wave functions that have been calculated 
within the framework of AMD
and discuss in particular the $nn$ correlations.


\subsection{Wave functions of $^6$He, $^8$He, reference states and intrinsic densities}
\begin{figure}[ht]
\epsfxsize=7 cm
\centerline{\epsffile{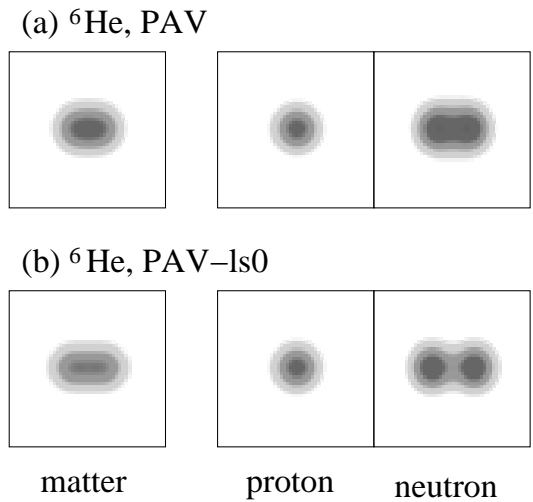}}
\vspace*{8pt}
\caption{\label{fig:he6-density}
One-body density distribution of the intrinsic wave functions of the reference state (PAV)
and the sample state (PAV-ls0) for $^6$He. Distribution of the
matter, proton and neutron density is shown left, middle and right, respectively.
}
\end{figure}
\begin{figure}[ht]
\epsfxsize=7 cm
\centerline{\epsffile{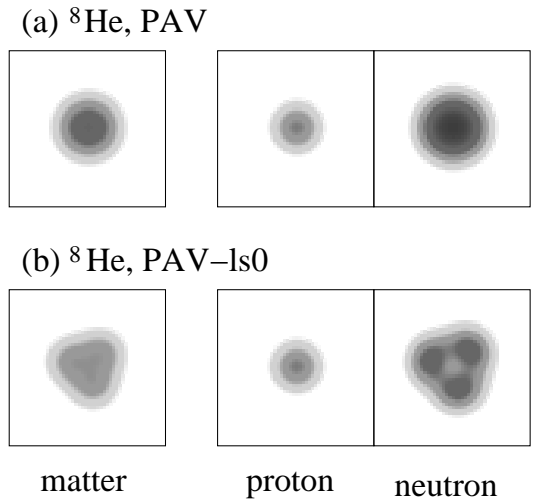}}
\vspace*{8pt}
\caption{\label{fig:he8-density}
Same as Fig.~\ref{fig:he6-density} but for $^8$He.}
\end{figure}
AMD calculations with multiconfiguration mixing (MC) for $^6$He and $^8$He were performed 
in Ref. \cite{KanadaEn'yo:2007ie}.
In the present paper we analyze the MC wave functions that were obtained in 
Ref. \cite{KanadaEn'yo:2007ie}
with the interaction parameter set "m56". 
It consists of the MV1 case(3) central force \cite{MV1} with parameters $m=0.56$, $b=h=0.15$ 
and the G3RS-type spin-orbit force \cite{LS} with the strengths $u_{I}=-u_{II}=2000$ MeV.
This interaction set gives a good reproduction of the ground state properties of 
$^6$He and $^8$He and the neutron-halo structure in $^6$He as well as a reasonable
description of the subsystem energies such as the $n$-$n$ scattering as shown 
in Ref. \cite{KanadaEn'yo:2007ie}.
Hereafter, we call this parameter set "m56-ls2000".

Each wave function of $^6$He and $^8$He is expressed by a linear combination of 
$k_{\rm max}=72$ parity and total-angular-momentum projected AMD configurations,
see Eq. (\ref{eq:confmix}).
In previous multiconfiguration AMD calculations for $^8$He, a second $0^+$ state with 
a well developed $^4$He+$2n$+$2n$ cluster structure was suggested,
though there is no experimental data for the existence of this excited $0^+$ state yet.
We investigate the two-neutron densities for the ground states of $^6$He$(0^+_1)$ 
and $^8$He$(0^+_1)$, and also that for the excited $^8$He$(0^+_2)$ state.

As reference states with less correlations, we prepare AMD wave functions 
by minimizing the energy of a single Slater determinant
$\Phi_{\rm AMD}({\bf Z})$ without any projection or constraint. 
After the variation this intrinsic state is projected on $J^\pi=0^+$.
This procedure is denoted by PAV (projection after variation).
The same interaction m56-ls2000 is used.

In order to study the effects of the spin-orbit interaction on the
two-neutron correlations
we also create sample states with
PAV calculations without the spin-orbit force,
$m=0.56$, $b=h=0.15$ and $u_{I}=-u_{II}=0$ MeV (m56-ls0). 
By switching off the spin-orbit force,
spin-zero neutron pairs are favored.
In the following we label the PAV wave functions of the reference states 
by "PAV" and those obtained with no spin-orbit force by "PAV-ls0" (see Table \ref{tab:intpara}). 

Let us briefly explain features of the intrinsic structure of the reference 
and sample states.
In Fig.~\ref{fig:he6-density}, one-body density distributions of the intrinsic 
reference state
$\Phi_{\rm AMD}({\bf Z})$ before projection (denoted by PAV)
and those of the sample state with no spin-orbit force (denoted by PAV-ls0) are illustrated
for $^6$He. 
The intrinsic proton density is spherical, as spin up and down protons fill the
$s$-shell.
The dumbbell shape of the intrinsic neutron density 
of the sample state (PAV-ls0) indicates two pairs of spin up and down neutrons
(Fig.~\ref{fig:he6-density}(b)), while in the reference state 
(PAV), due to the spin-orbit force, the spin-zero neutron pairs are less separated 
and the $p_{3/2}$ components increase (Fig.~\ref{fig:he6-density}(a)).

These effects are even more pronounced in the case of $^8$He displayed in
Fig.~\ref{fig:he8-density}.
In the intrinsic wave function of the reference state (PAV), 
one finds that
the 6 neutrons fill the $s_{1/2}$- and the $p_{3/2}$-shells
and thus form a spherical distribution, see Fig.~\ref{fig:he8-density}(a).
This intrinsic wave function is already a good approximation to a $J^\pi=0^+$ state,
and hence the PAV reference state for $^8$He is a single Slater determinant 
even
after $J^\pi$-projection. 
Therefore, we can regard this reference state as an uncorrelated state.

On the other hand, the intrinsic density of $^8$He in the sample state 
with no spin-orbit force (PAV-ls0) shows a triangular structure in the neutron density
which indicates the three localized pairs of spin up and down neutrons
(Fig.~\ref{fig:he8-density}(b)). 
After projection on $J^\pi=0$ all one-body densities will be
spherical but
the sample state contains this special kind of two-neutron correlations
which will show up in the two-body density.
\begin{table}[ht]
\caption{ \label{tab:intpara} Adopted parameter sets of the effective nuclear force
and labels of the wave functions for $^6$He, $^8$He and $^{12}$C. 
"MC" indicates the superposition of 
multiconfiguration AMD wave functions. "PAV" and "PAV-ls0" correspond to 
total angular momentum and parity projection after variation calculations
for the reference states and the sample states, respectively. The details are described in text. 
}
\begin{center}
\begin{tabular}{c|ccc}
\hline
  set & central & spin-orbit &  \\
&  MV1 case (3)  & G3RS &  \\ 
\hline
 m56-ls2000 \cite{KanadaEn'yo:2007ie} & $m=0.56$, $b=h=0.15$ & $u_{I}=-u_{II}=2000$ MeV & \\
 m56-ls0 &  $m=0.56$, $b=h=0.15$ & $u_{I}=-u_{II}=0$ MeV & \\
 m62-ls3000 \cite{KanadaEn'yo:2006ze} & $m=0.62$, $b=h=0$ & $u_{I}=-u_{II}=3000$ MeV & \\
 m62-ls0 &  $m=0.62$, $b=h=0$ & $u_{I}=-u_{II}=0$ MeV & \\
\hline
 &  multiconfigurations & \multicolumn{2}{c}{PAV calculations} \\
 & \multicolumn{3}{c}{label of the wave functions } \\
nucleus & MC & PAV & PAV-ls0 \\
   &  & (reference states) & (sample states) \\
\hline
$^6$He &  m56-ls2000 & m56-ls2000 & m56-ls0 \\
$^8$He &  m56-ls2000 & m56-ls2000 & m56-ls0 \\
$^{12}$C &  m62-ls3000 & m62-ls3000 & m62-ls0 \\
\hline
\end{tabular}
\end{center}
\end{table}

\subsection{One-body density}

\begin{figure}[ht]
\epsfxsize=7 cm
\centerline{\epsffile{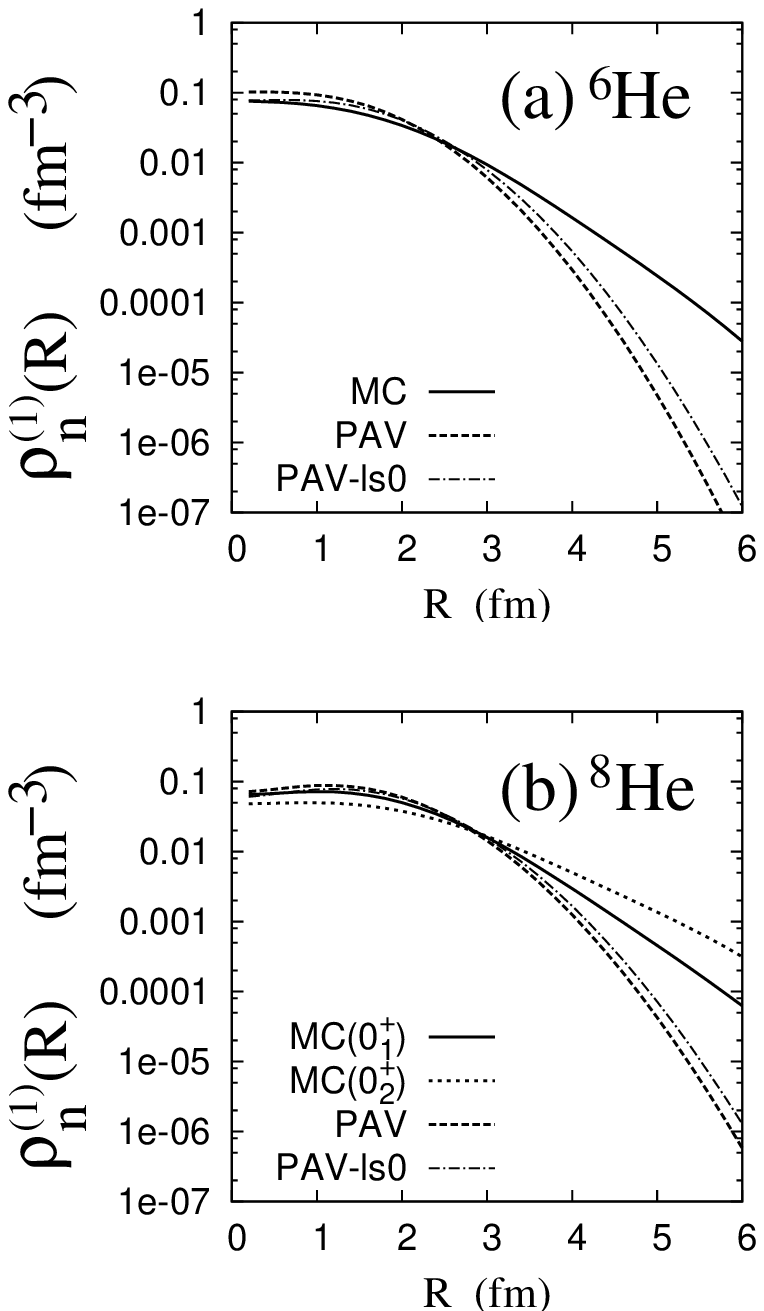}}
\vspace*{8pt}
\caption{\label{fig:he-dens1}
One-body neutron density $\rho_n^{(1)}(R)$ of the $^6$He$(0^+_1)$ and
$^8$He$(0^+_1)$ multiconfiguration states (MC), reference states (PAV)
and sample states (PAV-ls0), 
as well as that of the $^8$He$(0^+_2)$ multiconfiguration state (MC).
}
\end{figure}

The one-body densities of neutrons, $\rho^{(1)}_n(R)$, calculated from 
the MC wave functions for $^6$He($0^+_1$) and $^8$He($0^+_1$) are displayed 
in Fig.~\ref{fig:he-dens1} together with
those of the reference states (PAV) and of the sample states (PAV-ls0).
The density of the MC wave function for the excited state, $^8$He($0^+_2$),
is also shown.

\begin{figure}[ht]
\epsfxsize=7 cm
\centerline{\epsffile{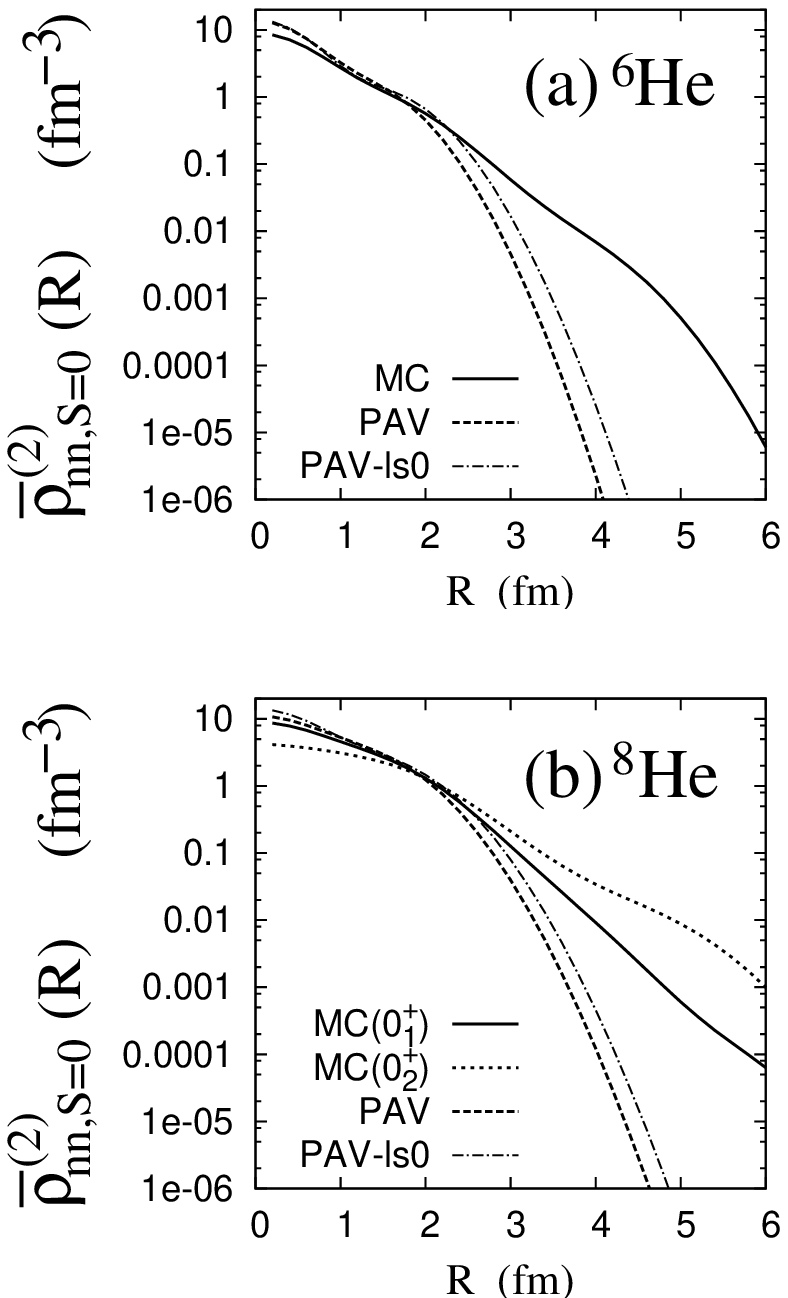}}
\vspace*{8pt}
\caption{\label{fig:he-dens2}
$S=0$ two-neutron c.m. density $\bar\rho_{nn,S=0}^{(2)}(R)$ of the $^6$He$(0^+_1)$
and $^8$He$(0^+_1)$ multiconfiguration states (MC), reference states (PAV)
and sample states (PAV-ls0), 
as well as that of the $^8$He$(0^+_2)$ multiconfiguration state (MC).
}
\end{figure}

In case of $^6$He$(0^+_1)$, 
the neutron density of the MC wave function shows in the outer region a low-density 
tail, the neutron-halo, while that 
of the reference state (PAV) has no noticeable tail. 
The sample state (PAV-ls0) has more neutron density at the nuclear surface but
the low-density tail does not extend so much as the MC one.
The reason is that in the MC state configurations are admixed in which
neutrons are further away from the core than in the reference state.

In the $^8$He($0^+_1$) MC state a similar behavior of the neutron tail 
is found. However, the difference from 
the reference state (PAV) is not as pronounced as in $^6$He.
On the other hand, the MC wave function  of $^8$He($0^+_2$) 
has a far reaching low-density tail in the neutron density.

One should keep in mind that the one-body density of a stationary
eigenstate of the Hamiltonian 
can only provide an indication of
many-body correlations 
but cannot prove their existence.
The same halo could in principle also be possible in a mean-field picture
where the last neutrons occupy weakly bound single-particle states.
Such a picture would of course not explain the Borromean behavior.
Therefore we explore in the following two-body densities
and study in which way correlations may affect two-body densities.


\subsection{Two-body density}

\subsubsection{Two-neutron c.m. density $\bar\rho^{(2)}_{nn,S=0}(R)$}

Equation (\ref{eq:rhobar}) defines the $S=0$ two-neutron c.m. density $\bar\rho^{(2)}_{nn,S=0}(R)$ 
to find a $S=0$ neutron pair with its c.m. position at $R$.
In Fig.~\ref{fig:he-dens2} the $S=0$ two-neutron c.m. density 
is displayed for $^6$He and $^8$He.
In the case of an uncorrelated
gas of neutrons one anticipates a more narrow distribution for the c.m. 
positions of pairs than that of the positions of individual neutrons because
particles on opposite sides of the nucleus contribute to c.m. positions at the center.
particles on opposite sides of the nucleus contribute to c.m. positions at the center.
This effect is nicely visible in Figs.~\ref{fig:he-dens1} and \ref{fig:he-dens2}. 
The one-body density of the uncorrelated $^6$He$(0^+_1)$ PAV state drops to the one percent level
of the central density around 3.6~fm while the two-body density does so at 
about 2.3~fm. Also for $^8$He$(0^+_1)$ case the difference is about 1.3~fm.
If on the other hand due the interaction between them the neutrons like
to form $S=0$ pairs that are preferentially close in distance the two-body
c.m. distribution will not be as narrow as in the uncorrelated case.
Therefore, $\bar\rho^{(2)}_{nn,S=0}(R)$
is a useful measure of the spin-zero $nn$ correlations.
An enhancement at the nuclear surface, for example, indicates that 
the halo contains preferentially correlated $S=0$ $nn$-pairs.

When compared with the reference states (PAV) and also the sample states (PAV-ls0), 
the two-neutron c.m. densities calculated with the MC wave functions for the $^6$He($0^+_1$) 
and $^8$He($0^+_1$) states are very large at the nuclear surface. 
In particular, $^6$He($0^+_1)$ shows a significant two-neutron c.m. density in the 
$R > 3$ fm region.
This suggests more enhanced $nn$ correlations at the surface of $^6$He than of $^8$He.
Even in the $^8$He($0^+_1)$ state the two-neutron c.m. density $\bar\rho^{(2)}_{nn,S=0}(R)$ 
of the MC state at $R\sim 4$~fm is by a factor $100$ larger than that of the reference state.
The excited $^8$He($0^+_2)$ state shows in the outer region
remarkably large probabilities for $S=0$ neutron pairs.

Compared with the results of the one-body density, 
the difference of the $S=0$ two-neutron c.m. densities between 
the MC wave functions and the PAV ones is 
in both cases striking.
This means that the $S=0$ two-neutron 
c.m. density is a much better indicator for $nn$ correlations than the one-neutron density
although both depend only on one variable, namely the distance from the center of the nucleus.

\subsubsection{Two-neutron probability densities $\rho^{(2)}_{nn}(R,r)$
and  $\rho^{(2)}_{nn,S=\{0,1\}}(R,r)$.}

The total two-neutron densities $\rho^{(2)}_{nn}(R,r)$ and their $S=0$ and $S=1$ parts
$\rho^{(2)}_{nn,S=\{0,1\}}(R,r)$  
are illustrated in Figs.~\ref{fig:he6-rho2s1} and ~\ref{fig:he8-rho2s1}
for the $0_1^+$ ground state of $^6$He and for the $0_1^+$ and $0_2^+$
states of $^8$He, respectively. 
The scale of the horizontal axis for $R$ is taken to be
$2\sqrt{A/(A-2)}$ times larger than that of the vertical axis for $r$.
In the limit of a simple uncorrelated state where two neutrons are moving 
in a $0s$ orbit around a core with mass $A-2$, the two-neutron density should 
be a function of $4R^2(A-2)/A+r^2$ and 
its contour lines become concentric circles in the 
($R,r$)-plane 
in this scaling. 
For example in Fig.~\ref{fig:he8-rho2s1} the contour lines of the total density belonging to the 
$^8$He PAV reference state are shown (most left panel in the second row). 
In the region far from the origin they look like concentric circles in the scaled 
($R,r$)-plane, indicating almost no correlation in the outer low-density region.

\subsubsection{$S=1$ two-neutron probability density $\rho^{(2)}_{nn,S=1}(R,r)$}

\begin{figure}[th]
\epsfxsize=16cm \epsffile{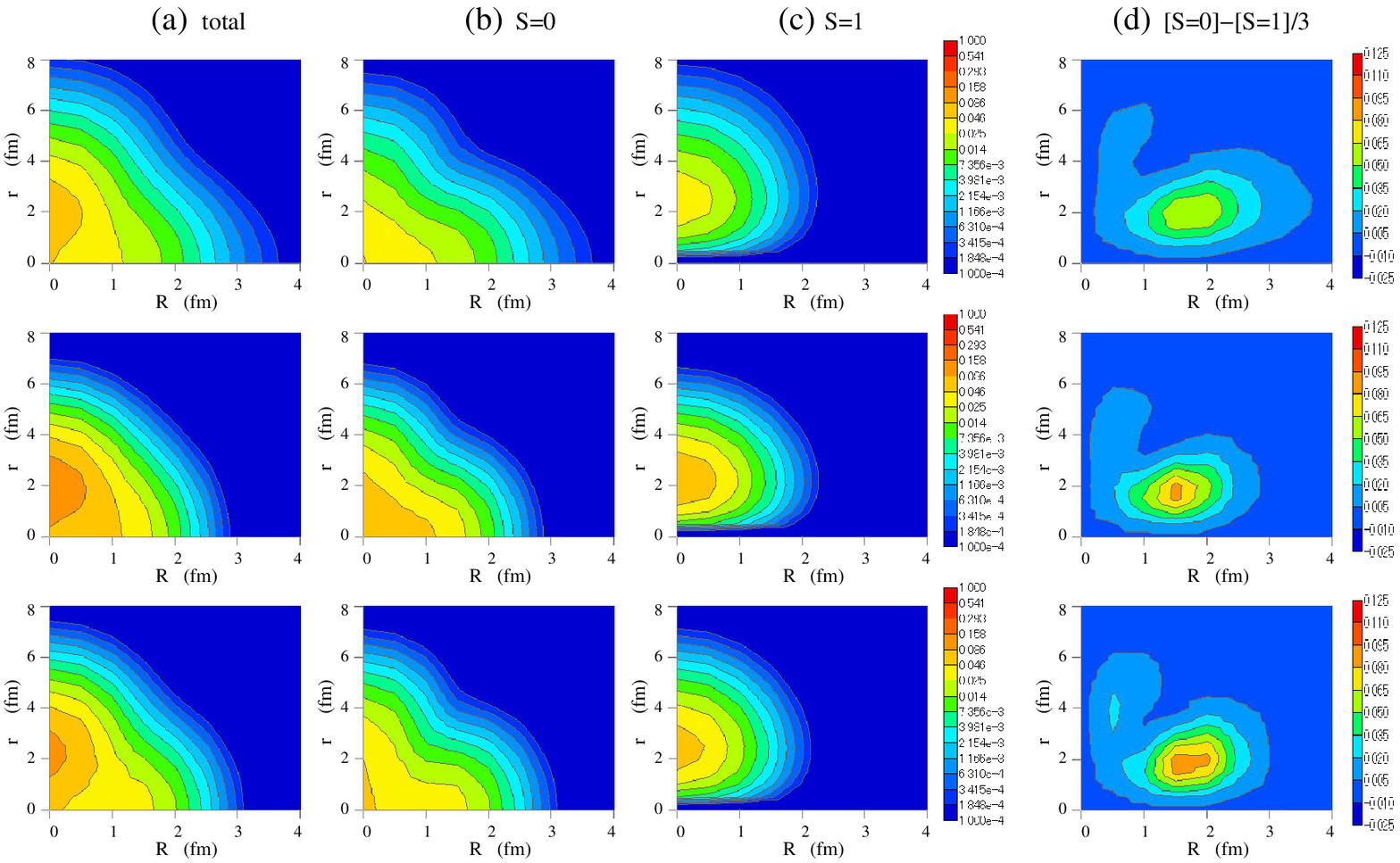}
\vspace*{8pt}
\caption{\label{fig:he6-rho2s1}
(Color online) Two-neutron densities of $^6$He$(0^+_1)$:
(a) total density $\rho_{nn}^{(2)}(R,r)$,  
(b) $S=0$ component $\rho_{nn,S=0}^{(2)}(R,r)$,  
(c) $S=1$ component $\rho_{nn,S=1}^{(2)}(R,r)$, 
(d) difference $r^2R^2(\rho^{(2)}_{nn,S=0}(R,r)-\rho^{(2)}_{nn,S=1}(R,r)/3)$.
First row for the MC state, second row for the PAV reference state and third row
for the PAV-ls0 sample state.
}
\end{figure}

\begin{figure}[th]
\epsfxsize=16cm \epsffile{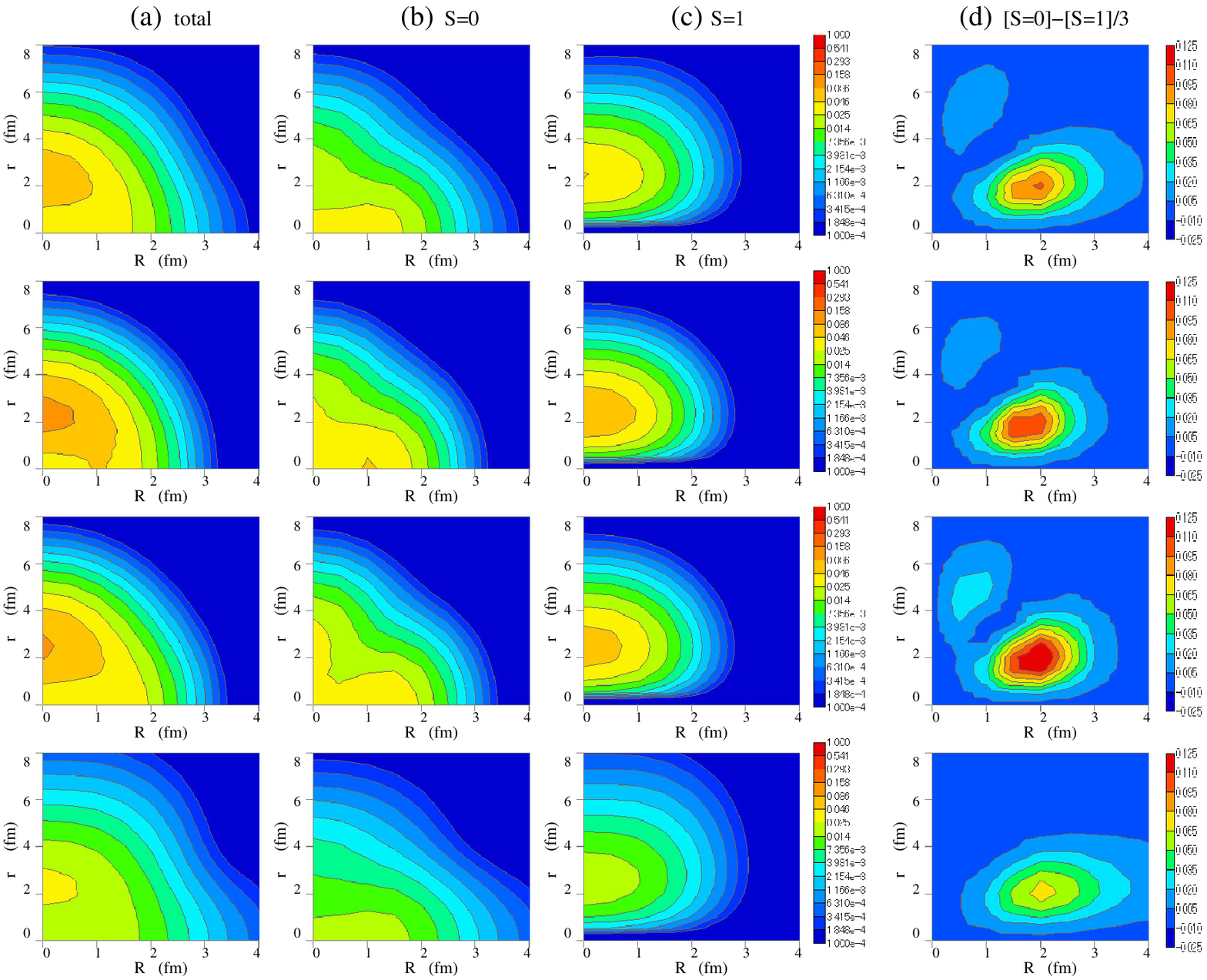}
\vspace*{8pt}
\caption{\label{fig:he8-rho2s1}
(Color online) Same as Fig.~\ref{fig:he6-rho2s1} but for $^8$He.
The forth row shows the densities of the MC $(0^+_2)$ excited state.
}
\end{figure}

A first correlation that originates from the Pauli principle, and thus is
regarded as trivial, can be seen in Figs.~\ref{fig:he6-rho2s1} and ~\ref{fig:he8-rho2s1}
at $r\approx 0$. While the the spin-zero component $\rho^{(2)}_{nn,S=0}(R,r)$
has large amplitude in the small $r$ region, the $S=1$
two-neutron density $\rho^{(2)}_{nn,S=1}(R,r)$ 
vanishes at $r=0$ and
concentrates in regions with large $r$.
This is easily understood because spatially odd (even) components of the relative motion
are automatically selected for $S=1$ ($S=0$) pairs in the antisymmetrized wave functions.


It is interesting to see in Fig.~\ref{fig:he6-rho2s1} that for all three states 
MC, PAV, and even for the unphysical PAV-ls0 state the $S=1$ two-neutron densities are 
rather similar for the $^6$He ground state. The same holds true for the ground state 
of $^8$He.
This can be seen in Fig.~\ref{fig:he8-rho2s1}, which shows the densities for $^8$He 
including the excited $0^+_2$ MC state.
Only the $S=1$ density of the excited $0^+_2$ MC state (bottom row) extends significantly 
over a wider range of distances $r$ between the neutron pair.

\subsubsection{$S=0$ two-neutron probability density $\rho^{(2)}_{nn,S=0}(R,r)$}

Let us investigate now the $S=0$ two-neutron density $\rho^{(2)}_{nn,S=0}(R,r)$ 
of the MC wave functions and compare with the results 
of the reference states (PAV) and the sample states (PAV-ls0).
The $S=0$ two-neutron densities of $^6$He
are plotted in the second column of Fig.~\ref{fig:he6-rho2s1}. 
A characteristic feature of the MC wave function is that 
its $S=0$ two-neutron density extends beyond $R > 3$ fm  without
broadening in the relative distance $r$ of the $nn$-pair. 
The occurrence of relatively small $r$ values at large c.m. distance $R$
indicates the presence of an extended long tail of a $S=0$ pair of two neutrons that  
are closer to each other than their c.m. distance from the core.
Such $nn$ correlations are not clearly visible in the 
reference state (PAV)
at large $R$.
Let us remind the reader that the MC wave function shows the enhancement of the 
$S=0$ two-neutron probability density 
already in the integrated
$\bar\rho^{(2)}_{nn,S=0}(R)$ at the nuclear surface $R >3$ fm,
see Fig.~\ref{fig:he-dens2}. In the two-dimensional plot it becomes clear
that the two neutrons reside in pairs with extensions considerably less
than only Pauli-correlated pairs.

On the other hand the two-neutron densities of the unphysical PAV-ls0 sample states
(third rows of  Fig.~\ref{fig:he6-rho2s1} and \ref{fig:he8-rho2s1}) 
indicate that in these states, due to the absence of the spin-orbit force,
the neutrons are grouped in $S=0$ pairs in the outer regions at large $R$.
These $nn$ correlations in the surface one can already anticipate from
the intrinsic one-body densities displayed in Fig.~\ref{fig:he6-density}(b) 
and \ref{fig:he8-density}(b).
But there one cannot judge in which regions the $nn$-pairs have predominantly 
$S=0$ or $S=1$. Looking in particular at the intrinsic one-body density 
of $^8$He in Fig.~\ref{fig:he8-density}(b) which shows three di-neutron clusters 
and has a minimum at $R=0$ one could even 
be misled and believe that the two-body density also should have a minimum 
at $R=0$. But the two-body density has actually a maximum at $R=0$.
The correct interpretation is that $nn$-pairs that are located vis-\`a-vis
from the center contribute at small c.m. $R$ and large relative $r$,
while those sitting next to each other contribute at large $R$ and small $r$.

Many-body states obtained by configuration mixing (MC) are able to represent
correlations beyond those residing in angular momentum projected intrinsic single 
Slater determinants. Especially the $0^+_1$ and $0^+_2$ MC states of $^8$He
in Fig.~\ref{fig:he8-rho2s1} (first and fourth row) show that the $S=0$
neutron pairs tend to be concentrated at small $r$ and to a lesser extent
at small $R$, quite in contrast to the uncorrelated PAV state or the
unphysical PAV-ls0 state.

From the above analysis of the two-neutron density, we can 
conclude that in the surface of the neutron-rich nuclei $^6$He and $^8$He 
the nuclear interaction induces $nn$ correlations of $S=0$ di-neutron character.

At this point we like to emphasize that neutrons are indistinguishable fermions 
and one cannot differentiate neutrons in the
$\alpha$-core from those in the valence orbits, therefore one 
cannot filter out the contribution of the valence pairs alone.
Let us consider the following spin structure
\begin{eqnarray}\label{eq:spin}
\nonumber
\big[s1\times s2\big]^{S=0}_{M=0}\times\big[s3 \times s4 \big]^{S=0}_{M=0}&=&
  \frac{1}{2}\ \big[s1 \times s3\big]^{S=0}_{M=0}\times \big[s2 \times s4 \big]^{S=0}_{M=0}\\
&+&\frac{1}{2}\ \big[s1 \times s3\big]^{S=1}_{M=1}\times \big[s2 \times s4 \big]^{S=1}_{M=-1}\\
\nonumber
&-&\frac{1}{2}\ \big[s1 \times s3\big]^{S=1}_{M=0}\times \big[s2 \times s4 \big]^{S=1}_{M=0}\\
\nonumber
&+&\frac{1}{2}\ \big[s1 \times s3]^{S=1}_{M=-1}\times \big[s2 \times s4 \big]^{S=1}_{M=1}\ ,
\end{eqnarray}
where $s1$, $s2$ denote spin-1/2 neutron states with spatial orbits in the core and $s3$, $s4$ 
spin-1/2 states with spatial orbits in the valence space. The core-core pair is
coupled to $S=0$ and so is the valence-valence pair. But as Eq.~(\ref{eq:spin})
shows this state can also be written as a superposition of 
a four-neutron state with two $S=0$ core-valence pairs
and three four-neutron states with two $S=1$ core-valence pairs.

In the extreme case where a system consists only of several $S=0$ neutron pairs
that are separated from each other, we can remove the inter-pair contributions
by taking the difference between $S=0$ and one third of $S=1$ pairs,
\begin{equation}\label{eq:tworho0-3}
r^2R^2\left(\rho^{(2)}_{nn,S=0}(R,r)-\frac{1}{3}\rho^{(2)}_{nn,S=1}(R,r)\right)\ .
\end{equation}
The factor $\frac{1}{3}$ takes into account that there are three times more $S=1$ 
inter-pair contributions
than $S=0$ ones. 
The differences calculated according to Eq.~(\ref{eq:tworho0-3})
are shown in Figs.~\ref{fig:he6-rho2s1} and \ref{fig:he8-rho2s1} in the most right columns.
As expected the contribution located at small $R$, which is supposed to come
mainly from inter-pair nucleons, is strongly reduced, while
the $nn$ correlations of the valence pairs
are seen in the case of the MC wave functions as an enhancement of the amplitude  
in the ($R >3$ fm, $r\approx 2$ fm) region when compared to the PAV state.
In the case of $^8$He (see Fig.~\ref{fig:he8-rho2s1}) this effect is very pronounced 
for the excited $0_2^+$ MC state.

\subsubsection{Two-neutron density $\rho^{(2)}_{nn,S=0}(R,r)$ at $r=0$}
\label{sec:prod}
\begin{figure}[b]
\centerline{\epsfxsize=8 cm\epsffile{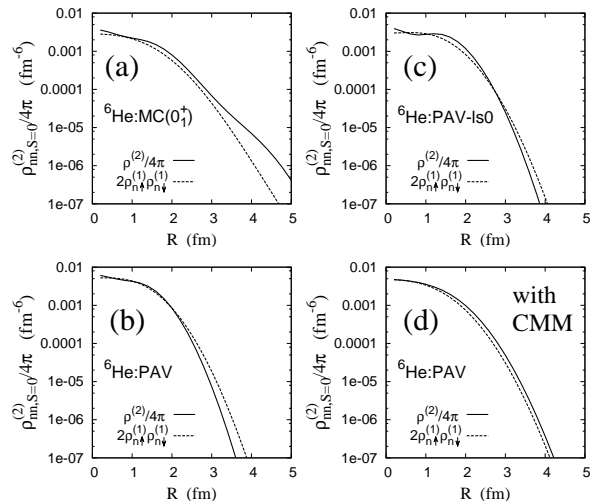}}
\caption{\label{fig:he6-rho-r0}
$\rho_{nn,S=0}^{(2)}(R,r=0)/4\pi$ and 
$2 \rho^{(1)}_{n\uparrow}(R) \rho^{(1)}_{n\downarrow}(R)$ for $^6$He.
(a), (b), and (c): Densities in the $^6$He$(0^+_1)$ multiconfiguration state (MC),
reference state (PAV), and the sample state (PAV-ls0), respectively, without the total c.m. motion.
(d): Densities in the reference state (PAV) with the total c.m. motion. 
}
\end{figure}

\begin{figure}[tt]
\centerline{\epsfxsize=8 cm\epsffile{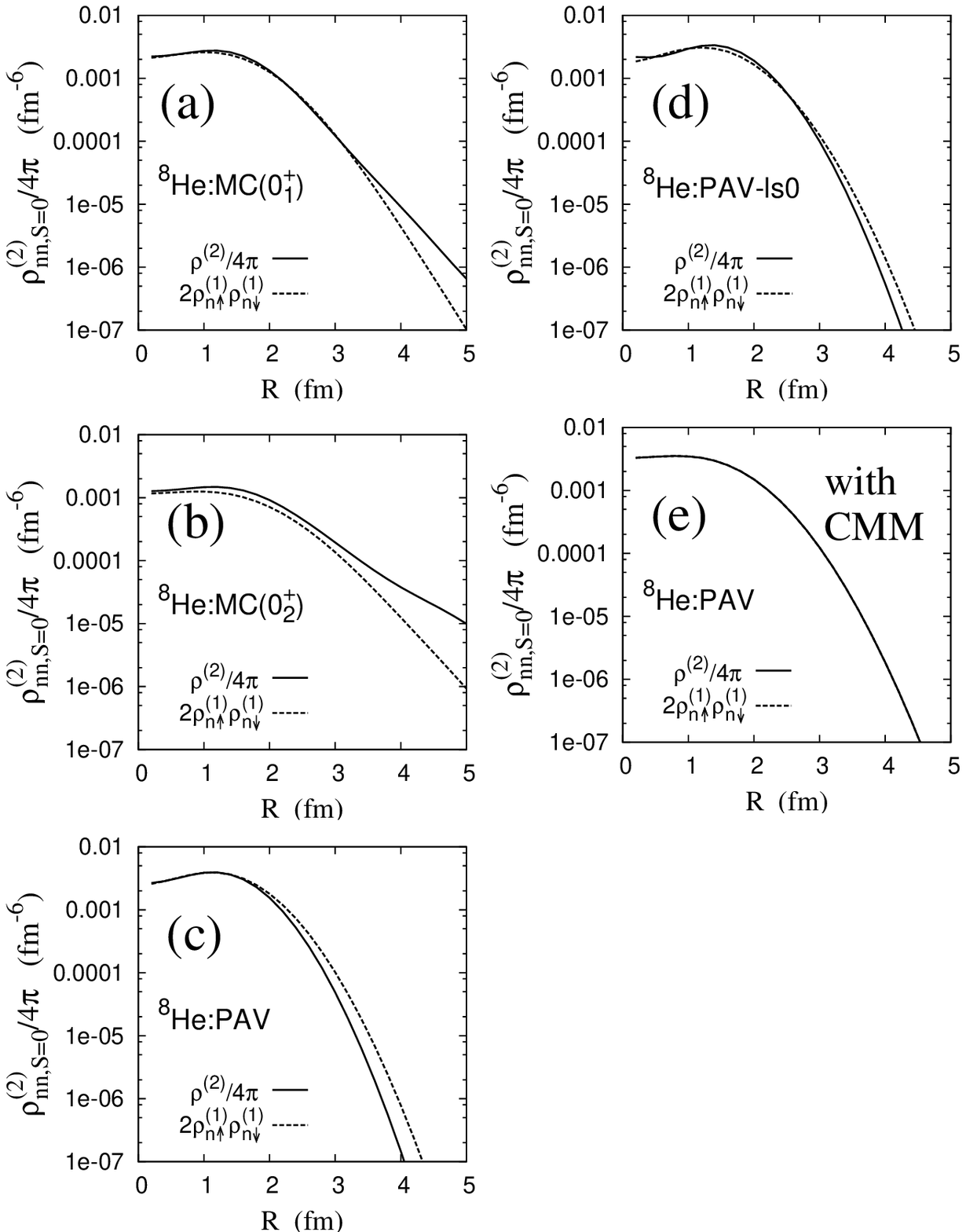}}
\caption{\label{fig:he8-rho-r0}
$\rho_{nn,S=0}^{(2)}(R,r=0)/4\pi$ and 
$2 \rho^{(1)}_{n\uparrow}(R) \rho^{(1)}_{n\downarrow}(R)$
for $^8$He.
(a), (b), (c), and (d): Densities in the $^8$He$(0^+_1)$ multiconfiguration state (MC), 
the $^8$He$(0^+_2)$ one,
the reference state (PAV), and in the sample state (PAV-ls0), respectively, without the total c.m. motion.
(e): Densities in the reference state (PAV) with the total c.m. motion. 
}
\end{figure}

As discussed above, the $S=0$ $nn$ correlations in the MC wave functions are characterized 
by a two-neutron density that extends toward large $R$ at small $r$ values. 
For a more quantitative discussion of the $S=0$ $nn$-pairs with strong spatial correlations, 
it is useful to  look at the two-neutron density at $r=0$, $\rho^{(2)}_{nn,S=0}(R,r=0)$.
The quantity $\rho^{(2)}_{nn,S=0}(R,r=0)/4\pi$ indicates the 
probability to find two neutrons at the 
same position ${\bf R}$, averaged over the orientation of ${\bf R}$ 
\begin{equation}
       \frac{1}{4\pi} \rho^{(2)}_{nn,S=0}(R,r=0)= 
      \frac{1}{4\pi} \int d\Omega_R \rho^{(2)}_{nn,S=0}({\bf R},{\bf r}=0) \ .
\end{equation}
Since the $S=1$ two-neutron density vanishes at $r=0$, 
$\rho^{(2)}_{nn,S=0}({\bf R},{\bf r}=0)$ equals to the total two-neutron density 
$\rho^{(2)}_{nn}({\bf R},{\bf r}=0)$. 

For a single Slater determinant the two-body density can be written as an antisymmetrized
product of one-body density matrices. It is easy to show that for ${\bf r}=0$, where only the
$S=0$ component contributes to the $nn$ density, this results in
\begin{equation}\label{eq:SD}
\rho^{(2),SD}_{nn,S=0}({\bf R},{\bf r}=0)=\rho^{(2)}_{nn}({\bf R},{\bf r}=0)=
    2 \rho^{(1)}_{n\uparrow}({\bf R}) \rho^{(1)}_{n\downarrow}({\bf R}) \ ,
\end{equation}
where $\rho^{(1)}_{n\uparrow}({\bf R})$ denotes the one-body spin-up neutron density, 
and analogue for spin-down. The relation holds only if the total c.m. motion is not eliminated.
For a $0^+$ state $\rho^{(2)}_{nn,S=0}({\bf R},{\bf r}=0)$ and 
$\rho^{(1)}_{n\uparrow}({\bf R})$
depend only on the absolute value $R$ so that
$\rho^{(2)}_{nn,S=0}({\bf R},{\bf r}=0)=\rho^{(2)}_{nn,S=0}(R,r=0)/4\pi$
and $\rho^{(1)}_{n\uparrow,\downarrow}({\bf R})=\rho^{(1)}_{n\uparrow,\downarrow}({R})$.
As a result, the relation
$\rho^{(2)}_{nn,S=0}(R,r=0)/4\pi=2 \rho^{(1)}_{n\uparrow}(R) \rho^{(1)}_{n\downarrow}(R)$
is satisfied for an uncorrelated $0^+$ state given by a single Slater determinant.
Therefore, comparing $\rho^{(2)}_{nn,S=0}(R,r=0)/4\pi$ with 
$2 \rho^{(1)}_{n\uparrow}(R) \rho^{(1)}_{n\downarrow}(R)$
gives another possibility to quantify
$nn$ correlations. Enhancement of  $\rho^{(2)}_{nn,S=0}(R,r=0)$
indicates that the many-body
wave function contains correlations of $S=0$ $nn$-pairs beyond the mean field level.
Here we should note that, when the total c.m. motion is properly removed from the  
uncorrelated $0^+$ state, the relation Eq.~(\ref{eq:SD})
is no longer satisfied because of the recoil effect. 

In Figs.~\ref{fig:he6-rho-r0} and \ref{fig:he8-rho-r0} the two-body densities at $r=0$,
$\rho^{(2)}_{nn,S=0}(R,r=0)/4\pi,$ are compared with the product of one-body densities
$2 \rho^{(1)}_{n\uparrow}(R) \rho^{(1)}_{n\downarrow}(R)$
of various many-body states for $^6$He and $^8$He, respectively.
The total c.m. motion is removed from the wave functions in the present
calculations as explained before.
For comparison, we also include 
the results for the PAV reference states with the total c.m. motion.
As seen in Fig.~\ref{fig:he8-rho-r0}(e) for the uncorrelated PAV reference state 
with the total c.m. motion 
for $^8$He, $\rho^{(2)}_{nn,S=0}(R, r=0)/4\pi$ agrees with 
$2 \rho^{(1)}_{n\uparrow}(R) \rho^{(1)}_{n\downarrow}(R)$
because the PAV wave function is equivalent to a closed $p_{3/2}$ neutron-shell 
configuration and can be written by a Slater determinant. 
After removing the total c.m. motion from the reference state, 
$\rho^{(2)}_{nn,S=0}(R, r=0)/4\pi$ is smaller 
than 
$2 \rho^{(1)}_{n\uparrow}(R) \rho^{(1)}_{n\downarrow}(R)$
in the surface region (Fig.~\ref{fig:he8-rho-r0}(c)),
because scaling down of $R$ due to the recoil effect 
is larger in two-body density than in one-body density in general.

Let us examine $\rho^{(2)}_{nn,S=0}(R, r=0)$ for the correlated states given 
by the MC wave functions.
The MC wave function for the $^6$He($0^+_1$) state shows remarkable enhancement of 
$\rho^{(2)}_{nn,S=0}(R,r=0)/4\pi$ 
at the surface compared with 
$2 \rho^{(1)}_{n\uparrow}(R) \rho^{(1)}_{n\downarrow}(R)$
(Figs~\ref{fig:he6-rho-r0}(a)).
This enhancement indicates clearly the 
higher correlations beyond mean field, which are incorporated by the MC calculation, i.e., 
the superposition of many $J^\pi$-projected Slater determinants.
By contrast, in the reference state (PAV) and the sample state (PAV-ls0),  
the surface tail of the two-neutron density $\rho^{(2)}_{nn,S=0}(R,r=0)/4\pi$ is smaller 
than that of the squared one-neutron density
just because of the recoil effect.
Also in $^8$He, the MC wave functions for the $^8$He($0^+_1$) and $^8$He($0^+_2$) states 
show enhancement of the two-neutron density  $\rho^{(2)}_{nn,S=0}(R,r=0)$ at the surface.
Compared with the results for the $^6$He($0^+_1$) state, it is found that 
the enhancement is less prominent in the $^8$He($0^+_1$) than in 
the $^6$He($0^+_1$). This means that the di-neutron correlations are weaker in the 
$^8$He ground state than in the $^6$He ground state.
On the other hand, the excited $^8$He($0^+_2$) shows the remarkably enhanced  
two-neutron density $\rho^{(2)}_{nn,S=0}(R, r=0)$ in the $R\ge 4$ fm region
because of the well developed $^4$He+$2n$+$2n$ structure.

\section{Results of $^{12}$C}\label{sec:12C}

Multi-configuration AMD calculations of $^{12}$C were performed 
in Refs.~\cite{KanadaEn'yo:2006ze,Enyo-c12} where the wave functions were obtained by 
variation after total-angular-momentum projection (VAP). It was shown that 
the AMD calculations successfully describe the ground state properties as well as 
various features of excited states with $3\alpha$ cluster structure.
In these studies of $^{12}$C, the $0^+_1$ state was found to be an admixture of 
the $p_{3/2}$ sub-shell closure and $SU(3)$-limit $3\alpha$ cluster components
while the $0^+_2$ state turned out to be a well-developed $3\alpha$-cluster state 
having the trend to form a gas like system of weakly interacting $\alpha$ particles.
Very similar results have been obtained in FMD studies \cite{Chernykh07,hoyle10} which
investigated in detail elastic and inelastic form factors.

Because a $S=0$ two-neutron pair is contained in each $\alpha$ cluster, 
the analysis of the $S=0$ two-neutron density is expected to 
be also helpful in identifying $\alpha$-cluster correlations in $^{12}$C.
In the following, we analyze the two-neutron densities of 
the $0^+_1$ and $0^+_2$ states of $^{12}$C and discuss the relation to the
$3\alpha$ cluster features.

\subsection{Wave functions of $^{12}$C} 

We use the wave functions of the $^{12}$C$(0^+_1)$ and $^{12}$C$(0^+_2)$ states 
that have been calculated in Ref.~\cite{KanadaEn'yo:2006ze}.
These MC wave functions are expressed as a linear combination of 
23 parity and total-angular-momentum projected AMD configurations 
which were obtained in VAP calculations.
The interaction parameter set "m62-ls3000" was used, see Table~\ref{tab:intpara}.

\begin{figure}[t]
\epsfxsize=7 cm
\centerline{\epsffile{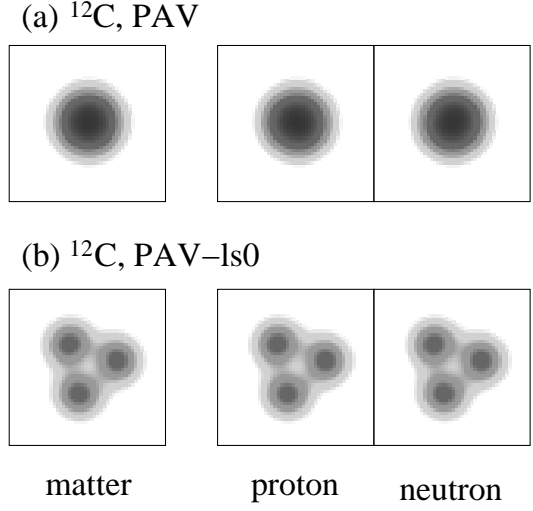}}
\vspace*{8pt}
\caption{\label{fig:c12-density}
One-body density distribution of the intrinsic wave functions of the reference state (PAV)
and the sample state (PAV-ls0) for $^{12}$C.}
\end{figure}
As a reference state with no or little correlations
we take the AMD state with minimum energy obtained by a PAV calculation 
using the same interaction "m62-ls3000".
Similar to the $^8$He case, the resulting $^{12}$C  reference state has intrinsically
an almost spherical shape as it is equivalent to $p_{3/2}$-shell closure,
see Fig.~\ref{fig:c12-density}(a).  
Therefore the $J^\pi$-projection for the reference state (PAV) changes little and the 
reference state is approximately a single Slater determinant 
which can be regarded as an uncorrelated state.

We also prepare the PAV-ls0 sample state by a PAV calculation with no spin-orbit force
by using interaction set m62-ls0. The intrinsic structure of 
the sample state (PAV-ls0) is also illustrated in Fig.~\ref{fig:c12-density}(b).
It shows  
a triangular configuration of  $3 \alpha$ clusters
because $\alpha$ clusters are energetically favored in absence of the spin-orbit force.

\begin{figure}[b]
\epsfxsize=5.5 cm
\centerline{\epsffile{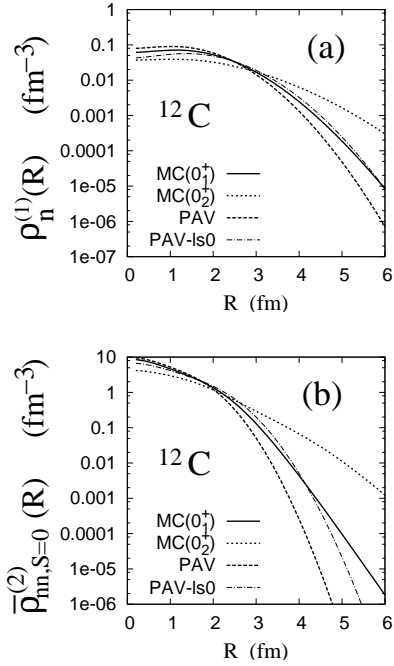}}
\vspace*{8pt}
\caption{\label{fig:c12-dens-fig}
(a) One-body neutron density $\rho_n^{(1)}(R)$ of the $^{12}$C$(0^+_1)$,
and $^{12}$C$(0^+_2)$ MC wave functions
as well as those of the reference states (PAV) and 
the sample state (PAV-ls0).
(b) $S=0$ two-neutron c.m. density $\bar\rho_{nn,S=0}^{(2)}(R)$
for the same states.
}
\end{figure}

\subsection{One-neutron densities and two-neutron c.m. densities}

In Fig.~\ref{fig:c12-dens-fig}(a) the one-body neutron densities, $\rho^{(1)}_n(R)$, 
of the MC wave functions are shown for the ground state $^{12}$C$(0^+_1)$ and 
the Hoyle state $^{12}$C$(0^+_2)$.
The one-body density of the ground state differs not so much from 
that of the uncorrelated reference state as in the cases of $^{6}$He and $^8$He.
On the other hand, the 
density of the
$^{12}$C$(0^+_2)$ Hoyle state is 
much lower in the interior and has a far out reaching tail.
By just looking at the one-body density one can not decide if the
many-body state is a shell model like state with individual 
nucleons moving in a shallow mean field or if, as is the case, 
the nucleons condense into $\alpha$-clusters which move in the outer
regions like a weakly interacting gas of $^4$He nuclei.

Let us discuss the $S=0$ two-neutron c.m. densities $\bar\rho_{nn,S=0}^{(2)}(R)$
of $^{12}$C states shown in Fig.~\ref{fig:c12-dens-fig}(b).
In the $^{12}$C($0^+_1$) MC ground state the enhancement of the two-neutron c.m. density 
at the surface is not so remarkable when compared with the reference state (PAV) and 
it is even less when compared with the sample state (PAV-ls0).
The reason is that
the ground state of $^{12}$C is  
well bound
with respect to the $3\alpha$ threshold, 
and therefore formation of $\alpha$ clusters 
is not expected at the surface.
This is in contrast to the cases of $^6$He and $^8$He which 
are loosely bound systems close to the two-neutron threshold.
On the other hand, 
the MC wave function for
the second $0^+$ state, 
which has an energy very close to the 3 $\alpha$ breakup threshold,
shows a well developed 3$\alpha$-cluster structure.
This in turn leads to an
enhanced $S=0$ two-neutron c.m. density in the large $R$ region.

\subsection{Two-neutron probability densities $\rho^{(2)}_{nn}(R,r)$ and 
$\rho^{(2)}_{nn,S=0,1}(R,r)$}

\begin{figure}[th]
\epsfxsize=16 cm
\centerline{\epsffile{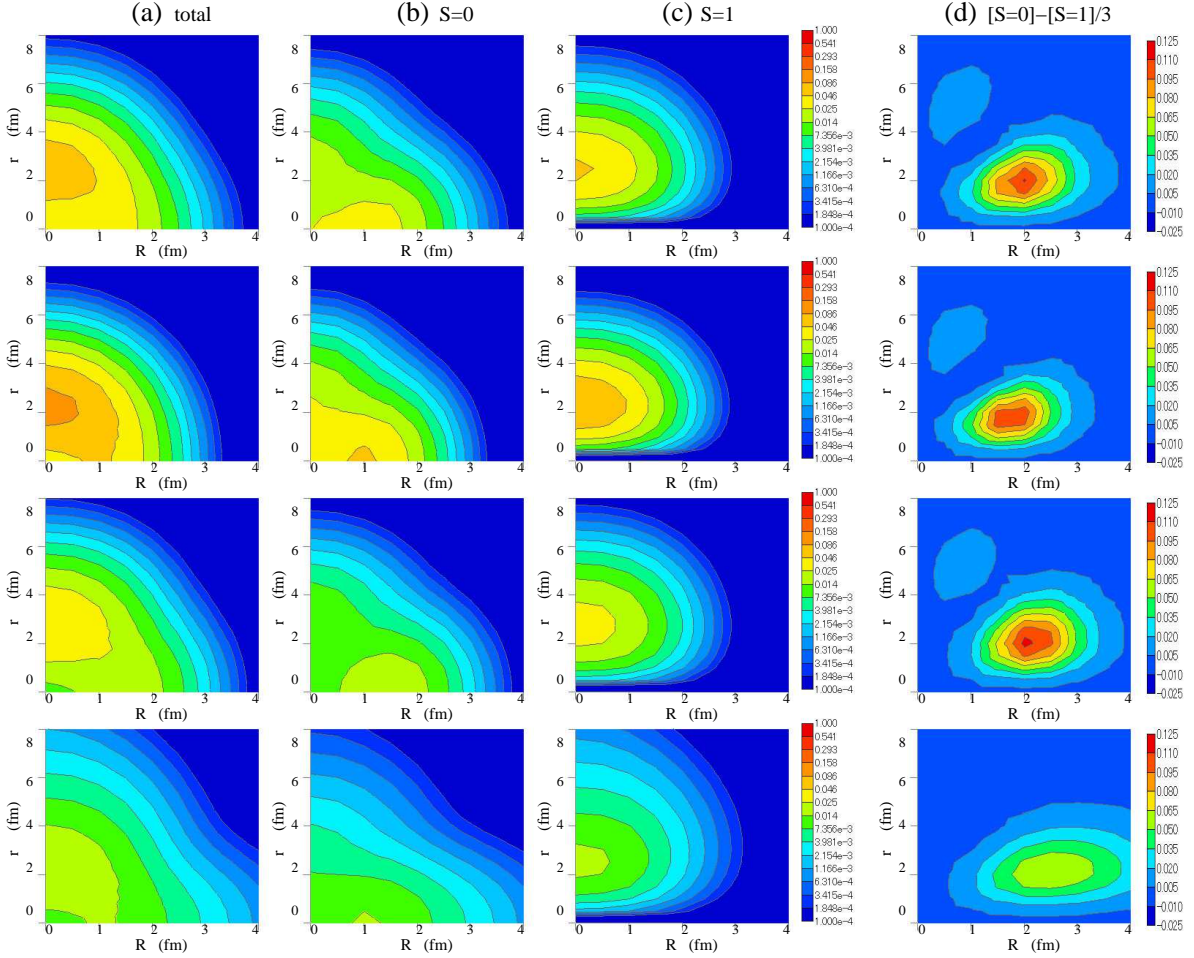}}
\vspace*{8pt}
\caption{\label{fig:c12-rho-f0} (Color online)
Two-neutron densities of $^{12}$C$(0^+_1)$:
(a) total density $\rho_{nn}^{(2)}(R,r)$,  
(b) $S=0$ component $\rho_{nn,S=0}^{(2)}(R,r)$,  
(c) $S=1$ component $\rho_{nn,S=1}^{(2)}(R,r)$, 
(d) difference $r^2R^2(\rho^{(2)}_{nn,S=0}(R,r)-\rho^{(2)}_{nn,S=1}(R,r)/3)$.
First row for the MC state, second row for the PAV reference state and third row
for the PAV-ls0 sample state. The forth row shows the densities of the MC $^{12}$C$(0^+_2)$ 
excited state Hoyle state.
}
\end{figure}
The calculated densities for the various states of $^{12}$C are summarized
in Fig.~\ref{fig:c12-rho-f0}. 
The two-neutron densities for the $^{12}$C PAV reference state are 
quite similar to those for the $^{8}$He PAV reference state because both
states have the neutron configuration of a $p_{3/2}$-shell closure. 
For example, in the total density for the $^{12}$C PAV reference state, 
the contour lines in the scaled $(R,r)$-plane look like concentric circles 
in the region far from the origin, indicating no correlations 
in the outer low-density region. As already discussed in the study of He isotopes, 
we can see the effect of two-body correlations, particularly, in $S=0$ two-neutron densities 
and also in the difference $r^2R^2(\rho^{(2)}_{nn,S=0}(R,r)-\rho^{(2)}_{nn,S=1}(R,r)/3)$
shown in the columns (b) and (d) in Fig.~\ref{fig:c12-rho-f0}.
Comparing the $S=0$ two-neutron densities of the
three wave functions for the $^{12}$C$(0^+_1)$ MC state, the PAV reference state and
the PAV-ls0 sample state, it is found that the spin-zero $nn$ correlations in the 
MC state are slightly 
stronger than those of the uncorrelated reference state (PAV), 
but weaker than those of the sample state (PAV-ls0).
This is consistent with the previous $^{12}$C study which suggested 
that the $^{12}$C ground state is an admixture of 
a shell model component with a
$p_{3/2}$ sub-shell closure and $3\alpha$-cluster components.

The second $0^+$ state of the MC result (last row of Fig.~\ref{fig:c12-rho-f0}) 
shows an enhanced amplitude of the $S=0$ two-neutron density for $R >3$ fm 
and below $r\approx 3$ fm. This is also clearly seen in the
difference Eq.~(\ref{eq:tworho0-3})  displayed in the last column of Fig.~\ref{fig:c12-rho-f0}.

\begin{figure}[th]
\centerline{\epsfxsize=8 cm\epsffile{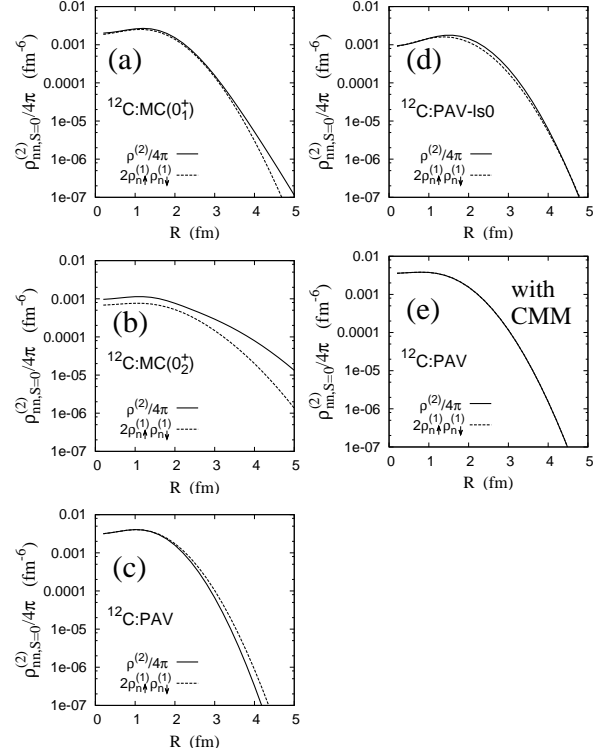}}
\caption{\label{fig:c12-rho-r0}
$\rho_{nn,S=0}^{(2)}(R,r=0)/4\pi$ and 
$2 \rho^{(1)}_{n\uparrow}(R) \rho^{(1)}_{n\downarrow}(R)$
for $^{12}$C.
(a), (b), (c), and (d): Densities in the $^{12}$C$(0^+_1)$ multiconfiguration state (MC), 
the $^{12}$C$(0^+_2)$ one,
the reference state (PAV), and the sample state (PAV-ls0), respectively, 
without the total c.m. motion.
(e): Densities in the reference state (PAV) with the total c.m. motion.}
\end{figure}

For a more quantitative discussion we compare in Fig.~\ref{fig:c12-rho-r0} the 
two-neutron density at $r=0$, $\rho^{(2)}_{nn,S=0}(R,r=0)/4\pi$, with 
the product of one-neutron densities, 
$2 \rho^{(1)}_{n\uparrow}(R) \rho^{(1)}_{n\downarrow}(R)$.
The features of the two-neutron density are qualitatively similar to those in $^8$He.
Namely, the MC wave functions show that the two-neutron density $\rho^{(2)}_{nn,S=0}(R,r=0)$
in the ground state of $^{12}$C is slightly enhanced at the surface. 
However the excited state, $^{12}$C$(0^+_2)$, shows a remarkable enhancement 
at all values of $R$ even in the center of the nucleus.
The reason lies in
the well-developed $3\alpha$ cluster structure
and the fact that each $\alpha$ cluster houses an $S=0$ $nn$-pair. Thus
the enhancement of $\rho^{(2)}_{nn,S=0}(R,r=0)$ can also be observed
in states with strong $\alpha$-cluster correlations.

\newpage

\section{Summary and outlook}\label{sec:summary}

Two-neutron correlations in $^6$He and $^8$He are investigated by analyzing the
two-body density of microscopic many-body wave functions obtained by 
antisymmetrized molecular dynamics (AMD).
In order to visualize non-trivial spatial correlations, that are induced by the neutron-neutron 
interaction, the two-neutron density is calculated as function of the distance, $r$,
and mean c.m. position, $R$, of the $nn$ pair.
Results from correlated AMD wave functions are
compared to those of uncorrelated (or less correlated) wave functions. These
reference states are taken to be single Slater determinants because 
they represent independent fermions, and correlations 
induced  by the Pauli principle are regarded as trivial. 
We find characteristic non-trivial two-neutron correlations in the $S=0$ channel as 
an enhancement of the two-neutron density $\rho^{(2)}_{nn,S=0}(R,r)$ toward 
large $R$ at small $r$ values. 
These two-neutron correlations are weaker in the ground state of $^8$He than in $^6$He
and are particularly pronounced at the surface of the excited $^8$He$(0^+_2)$ state.
It is also found that superpositions of many angular momentum and parity projected
Slater determinants are essential to incorporate the di-neutron correlations.

To see how $nn$ correlations are reflecting $\alpha$ cluster structures
the $0^+_1$ ground state and the first excited $0^+_2$ state (Hoyle state) of $^{12}$C 
are also investigated.  
As the Hoyle state consists to a large extent of three loosely bound 
$\alpha$-particles at large distances from the center it contains spatially correlated $nn$ pairs 
which are not present in the ground state. 
In this case the cause for finding neutron pairs spatially close is the property
of the nucleon-nucleon interaction to bind two protons and two neutrons  
particularly well so that strong 4-body correlations in form of $\alpha$ clusters
are developed. Also these correlations are nicely visualized. 

Thus, the two-neutron density is found to be a good probe to identify two-neutron 
correlations. In particular, the comparison of the two-neutron density with 
the squared one-neutron density, both calculated from the same many-body state, 
is useful for a quantitative discussion of the two-neutron correlations.

Although we do not expect that the general nature of the $nn$ correlations will be altered
one should investigate if realistic effective interactions which reproduce
the experimental phase shifts, like the ones obtained in the Unitary Correlation 
Operator Method (UCOM) \cite{ucom10,ucom98,ucom03,ucom04} and successfully used in FMD calculations 
\cite{neff01}, 
give the same results as the more phenomenological effective potentials used here. 

The other question is if large-scale shell model Hilbert spaces can equally well
represent surface di-neutron correlations as the AMD states.

It would also be interesting to see in how far di-neutron correlations exist
in heavier neutron-rich nuclei which possess a neutron skin but not necessarily a halo.
Is there also a transition from mean-field dominated to correlation dominated dynamics
as seen in lighter nuclei?

\section*{Acknowledgments}
The discussions during the EMMI-EFES Workshop held at GSI in February 2009, and the YITP 
workshop held at YITP in May 2009 were helpful to initiate this work.
The computational calculations of this work were performed by using the
supercomputers at YITP and done in Supercomputer Projects 
of High Energy Accelerator Research Organization (KEK).
This work was supported by Grant-in-Aid for Scientific Research from Japan Society for the 
Promotion of Science (JSPS).
It was also supported by the Grant-in-Aid for the Global COE Program "The Next Generation of Physics, 
Spun from Universality and Emergence" from the Ministry of Education, Culture, Sports, Science and
Technology (MEXT) of Japan. Part of this work has been done while one of us (H.F.)
was a visiting professor at the YITP.


\end{document}